\documentclass[12pt]{article}

\usepackage{graphicx}
\usepackage{amsmath}
\usepackage{amssymb}
\usepackage{cite}
\usepackage{bm}

\makeatletter

\numberwithin{equation}{section}

\DeclareMathOperator{\Det}{Det}
\DeclareMathOperator{\Ai}{Ai}
\DeclareMathOperator{\Res}{Res}

\newcounter{aff}

\makeatother

\begin{document}
\begin{titlepage}
\begin{flushright}
{\footnotesize NITEP 26, OCU-PHYS 508, USTC-ICTS-19-22}
\end{flushright}
\begin{center} 
{\Large\bf Quantum Mirror Map for Del Pezzo Geometries}\\
\bigskip\bigskip
{\large
Tomohiro Furukawa\,\footnote{\tt furukawa@sci.osaka-cu.ac.jp},
\quad
Sanefumi Moriyama\,\footnote{\tt moriyama@sci.osaka-cu.ac.jp},
\quad
Yuji Sugimoto\,\footnote{\tt sugimoto@ustc.edu.cn}}\\
\bigskip
${}^{*\dagger}$\,{\it\small Department of Physics, Graduate School of Science, Osaka City University,}\\
${}^\dagger$\,{\it\small Nambu Yoichiro Institute of Theoretical and Experimental Physics (NITEP),}\\
${}^\dagger$\,{\it\small Osaka City University Advanced Mathematical Institute (OCAMI),}\\
{\it\small Sumiyoshi-ku, Osaka 558-8585, Japan}\\[6pt]
${}^\ddagger$\,{\it\small Interdisciplinary Center for Theoretical Study, University of Science and Technology of China,}\\
{\it\small Hefei, Anhui 230026, China}
\end{center}

\begin{abstract}
Mirror maps play an important role in studying supersymmetric gauge theories.
In these theories the dynamics is often encoded in an algebraic curve where two sets of periods enjoy the symplectic structure.
The A-periods contribute to redefinitions of chemical potentials known as mirror maps.
Using the quantization of the $D_5$ del Pezzo geometry, which enjoys the symmetry of the $D_5$ Weyl group, we are able to identify clearly the group-theoretical structure and the multi-covering structure for the mirror map.
With the structures, we can apply the mirror map to superconformal Chern-Simons theories describing the worldvolume of multiple M2-branes on various backgrounds, where we find that the redefinition of the chemical potential is obtained directly from the mirror map.
Besides, we have interesting observations for the mirror map:
The representations appearing in the quantum mirror map are the same as those appearing in the BPS indices except for the trivial case of degree 1 and the coefficients are all integers.
\end{abstract}
\end{titlepage}

\setcounter{footnote}{0}

\tableofcontents

\section{Introduction and summary}

Mirror map is one of the most fascinating subjects in recent studies of supersymmetric gauge theories.
Classically \cite{CDGP} it was known that the Yukawa couplings in the complex structure parameters are not renormalized by instanton effects while those in the K\"ahler class parameters are, where two sides are related by mirror maps for the mirror pair of Calabi-Yau manifolds.
Later it turns out that similar structures are omnipresent in various examples, including matrix models or supersymmetric gauge theories.

Typically in supersymmetric gauge theories the dynamics is encoded in a classical algebraic curve where two sets of period integrals (called A-periods and B-periods) obtained by integrating the meromorphic one-form along two sets of cycles (called A-cycles and B-cycles correspondingly) enjoy the symplectic structure.
Depending on the physical system, the two sets of periods give respectively physical variables such as gluino condensations or chemical potentials and derivatives of the physical quantity characteristic of the system such as prepotential or free energy \cite{S,SW,ADKMV,NS,ACDKV}.

Since the derivatives of free energy characterizing the dynamics are obtained from the B-periods,\footnote{Usually, the B-periods are expressed in terms of the complex structure moduli and it is only after we reexpress them by the K\"ahler parameters using the A-periods that we encounter the positive integers characterizing the derivatives of free energy.
Hence in this sense, it is also common to regard them as obtained not only from the B-periods but the A-periods as well.
} the B-periods were studied extensively and it was found that they are determined by a set of positive integers called the Gopakumar-Vafa invariants \cite{GV,GV2} or their generalizations known as the BPS indices.
It turns out that these positive integers sometimes enjoy nice group-theoretical properties for the application to supersymmetric gauge theories.
With these properties, after identifying them as representations and specifying the unbroken subgroup preserved by the background, we can reconstruct the free energy for supersymmetric gauge theories without difficulty.

On the other hand, mirror maps (redefining physical variables such as chemical potentials) are obtained from the A-periods by integrating along the A-cycles.
Although the mirror maps serve half of the role in determining the dynamics, compared with the progress in the derivatives of free energy characterizing the dynamics directly, the progress in the mirror maps is only made gradually.
For example, although the mirror map is given in terms of the $A_1$ characters in \cite{HMMO} for the ${\mathbb P}^1\times{\mathbb P}^1$ case, the symmetry group is not large enough to convince ourselves of the group-theoretical structure.
Also it is desirable to confirm the multi-covering structure proposed in \cite{HMMO} with a larger symmetry group.
With insufficient understanding of the group-theoretical structure and the multi-covering structure for the A-periods, it is difficult to understand the redefinition of the chemical potentials for supersymmetric gauge theories.
Interestingly, it was proposed in \cite{AKV} that the multi-covering structure in the mirror map describes the BPS states in the presence of D-brane domain walls, which may be a clue to clarify the physical interpretation of the redefinition.

The supersymmetric gauge theories we have in mind for the application of the mirror map are three-dimensional ${\cal N}=3$ superconformal Chern-Simons gauge theories.
The most standard example is the ABJM theory, that is, the ${\cal N}=6$ superconformal Chern-Simons theory with gauge group U$(N)_k\times$U$(N)_{-k}$ (the subscripts $(k,-k)$ denoting the Chern-Simons levels) and two pairs of bifundamental matters, describing the worldvolume of $N$ coincident M2-branes on the target space ${\mathbb C}^4/{\mathbb Z}_k$ \cite{ABJM}.
Let us define the partition function of $N$ M2-branes on ${\mathbb C}^4/{\mathbb Z}_k$ as $Z_k(N)$ and move to the grand partition function
\begin{align}
\Xi_k(z)=\sum_{N=0}^\infty z^NZ_k(N),
\label{gcpf}
\end{align}
by regarding the rank $N$ of the ABJM theory as a particle number and introducing the dual fugacity $z$.
On one hand, we can rewrite the grand partition function into the Fredholm determinant \cite{MP}
\begin{align}
\Xi_k(z)=\Det(1+z\widehat H^{-1}),
\label{gpf}
\end{align}
of a quantum-mechanical spectral operator
\begin{align}
\widehat H=\widehat{\cal Q}\widehat{\cal P},\quad
\widehat{\cal Q}=\widehat Q^{\frac{1}{2}}+\widehat Q^{-\frac{1}{2}},\quad
\widehat{\cal P}=\widehat P^{\frac{1}{2}}+\widehat P^{-\frac{1}{2}},
\label{hhat}
\end{align}
which is reminiscent of the geometry ${\mathbb P}^1\times{\mathbb P}^1$ after change of variables.
On the other hand, with the expression of the Fredholm determinant, we can investigate the partition function systematically and find that the non-perturbative part is described by the free energy of refined topological strings on the same local ${\mathbb P}^1\times{\mathbb P}^1$ geometry \cite{AKMV,MPtop}.
Namely, aside from the perturbative part summing up to the Airy function \cite{DMP1,HKPT,FHM} reproducing the degrees of freedom $\log\Ai(N)\sim N^{\frac{3}{2}}$ well-known from the gravity analysis \cite{KT}, the non-perturbative part consists of worldsheet instanton effects \cite{DMP1}, membrane instanton effects \cite{DMP2,HMO2,CM} and their bound states \cite{HMO3} as well.
The non-perturbative part behaves nicely since, although coefficients of all the instanton effects are divergent at infinitely many values of $k$, the divergences are all cancelled among themselves \cite{HMO2}.
This cancellation can be understood much better if we redefine the chemical potential $\mu=\log z$ suitably into $\mu_\text{eff}=\log z_\text{eff}$ often called the effective chemical potential as in \cite{HMO3}
\begin{align}
\mu_\text{eff}=\mu+\frac{\pi^2k}{2}\sum_{\ell=1}^\infty a_\ell e^{-2\ell\mu},
\label{mueffredef}
\end{align}
with some coefficients $a_\ell$.
After the redefinition, all of the bound states are taken care of by the worldsheet instantons and the cancellations occur simply between the worldsheet instantons and the membrane instantons.
From the requirement of the cancellation we can further proceed to determine the whole function for the non-perturbative part and identify them as the free energy of refined topological strings \cite{HMMO}.

In particular, the free energy enjoys a multi-covering structure, where effects of lower degrees can appear in those of higher degrees.
Due to this structure the effects can be boiled down to the BPS indices $N^d_{j_\text{L},j_\text{R}}$ discussed previously.
Depending on degree $d$ and spins $(j_\text{L},j_\text{R})$ of the spacetime $\text{SO}(4)=\text{SU}(2)\times\text{SU}(2)$ the total BPS indices $N^d_{j_\text{L},j_\text{R}}$ were computed in \cite{HKP} for several geometries.
After assigning various K\"ahler parameters ${\bm T}$ for the geometry, the total BPS indices $N^d_{j_\text{L},j_\text{R}}$ are decomposed as $N^{\bm d}_{j_\text{L},j_\text{R}}$ by various combinations of degrees ${\bm d}$ (with $|{\bm d}|=d$) corresponding to the K\"ahler parameters ${\bm T}$ \cite{MNN}.
It was known that, for curves with group-theoretical structures such as curves of genus one known as del Pezzo geometries, the BPS indices inherit the group-theoretical structures of the curves.
Namely, the BPS indices form representations of the group and are specified by multiplicities of the irreducible representations $n^{d,{\bf R}}_{j_\text{L},j_\text{R}}$ \cite{MNY}.
Furthermore, for a certain fixed background geometry determined by the K\"ahler parameters, the BPS indices are split as the decompositions of representations into an unbroken subgroup preserved by the background \cite{KMN,KM}.
Thus, with the group-theoretical structure and the multi-covering structure clarified, the free energy is determined completely by the BPS indices.
For the description to work, however, the redefinition of the chemical potential is crucial.

Following the idea of the periods as we have explained generally at the beginning, it is not difficult to imagine that, the BPS indices are obtained from the B-period, while the redefinition of the chemical potential $\mu=\log z$ into the effective one $\mu_\text{eff}=\log z_\text{eff}$ is encoded in the A-period, which is in fact the case \cite{HMMO}.
Note, however, that to describe these supersymmetric gauge theories the algebraic curve in \eqref{hhat} is quantized \cite{NS,MM}, which means that the canonical variables $\widehat Q=e^{\widehat q}$ and $\widehat P=e^{\widehat p}$ describing the curve obey the commutation relation $[\widehat q,\widehat p]=i\hbar$ as in quantum mechanics.
Accordingly, the two sets of classical periods obtained by integrating the meromorphic one-form along cycles are promoted to quantum periods which are defined using the wave function for the Hamiltonian $\widehat H$ \cite{ACDKV}.
Hence, we are naturally led to the concept of the quantum mirror map and the quantum-corrected free energy \cite{NS}.

Although the group-theoretical structure and the multi-covering structure for the B-period were studied carefully previously, not so much was known for the quantum mirror map obtained from the A-period.
In \cite{HMMO} these structures of the quantum mirror map were proposed and studied, though it is desirable to study them from an example with a larger symmetry group to explore general structures systematically.

There are several generalizations for the ABJM theory which share the interpretation as the worldvolume theory of M2-branes and enjoy larger symmetry groups.
In this paper we concentrate especially on two ${\cal N}=4$ superconformal Chern-Simons theories connected \cite{MNN,KM} by the Hanany-Witten transitions \cite{HW}.
One of them is the ${\cal N}=4$ Chern-Simons theory with gauge group U$(N)_k\times$U$(N)_0\times$U$(N)_{-k}\times$U$(N)_0$ and bifundamental matters connecting subsequent group factors (called $(2,2)$ model after the powers appearing in the spectral operator $\widehat H^{(2,2)}=\widehat{\cal Q}^2\widehat{\cal P}^2$ \cite{MN1,MN3}), while the other is that with gauge group U$(N)_k\times$U$(N)_{-k}\times$U$(N)_k\times$U$(N)_{-k}$ and bifundamental matters (called $(1,1,1,1)$ model after $\widehat H^{(1,1,1,1)}=\widehat{\cal Q}\widehat{\cal P}\widehat{\cal Q}\widehat{\cal P}$ \cite{HM}).
Unlike the ABJM case where the symmetry group is only $A_1$, this time the symmetry group is $D_5$ and it is possible to study the quantum mirror map more systematically.

In this paper we study the quantum mirror map for the $D_5$ curve carefully, where we can identify the group-theoretical structure and the multi-covering structure explicitly.
We stress that the coefficients are only positive integers when we introduce extra signs correctly in the multi-covering structure.
Moreover, we observe that the representations appearing in the quantum mirror map are the same as those appearing in the BPS indices for each degree except for the trivial case of degree 1.

With these structures we are able to apply the mirror map to the superconformal Chern-Simons theories.
Namely, by specifying the parameters of the $D_5$ curve correctly, we can reproduce the effective chemical potential for the $(2,2)$ model and the $(1,1,1,1)$ model correctly.
As explained previously, the redefinition of the chemical potential takes care of all the bound states with the worldsheet instantons and reduces the complicated cancellation of divergences into the cancellation purely between the worldsheet instantons and the membrane instantons.
Considering these results, we are naturally led to the physical picture that the worldsheet instantons are dressed by the bound states and enhanced into ``effective'' worldsheet instantons containing an infinite tower of states.
After reducing the redefinition of the chemical potential into a series of positive integers, it is natural to regard these integers as the number of BPS states associated to the worldsheet instantons.
This viewpoint is reminiscent of the interpretation that the classical cousins of the integers count the number of BPS states in the presence of D-brane domain walls \cite{AKV}.

The organization of this paper is as follows.
In the next section, we first recapitulate the setup for the quantum mirror maps especially focusing on the case of the $D_5$ quantum curve.
Then in section \ref{QMPcharacter} we head for the study of the quantum mirror map for the $D_5$ curve and observe clearly the group-theoretical structure and the multi-covering structure.
After that in section \ref{mm} we turn to the analysis of gauge theories and find that our quantum mirror map reproduces the effective chemical potential correctly.
Finally, we conclude with discussions on further directions in section \ref{conclude}.

\section{Quantum mirror map}

\begin{figure}[!t]
\centering\includegraphics[scale=0.6,angle=-90]{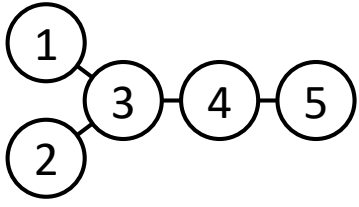}
\caption{Dynkin diagram of the $D_5$ algebra.}
\label{dynkin}
\end{figure}

To explain the symmetry breaking pattern for rank deformations of the $(2,2)$ model and the $(1,1,1,1)$ model \cite{MN3,MNN,MNY}, quantum curves were introduced and the $D_5$ quantum curve was studied carefully in \cite{KMN,KM}.
In this section we recapitulate the setup for the quantum curves following \cite{KMN} and the expression for the quantum mirror maps following \cite{ACDKV,HMMO}.
We define quantum curves to be Hamiltonians given by the canonical variables $\widehat Q=e^{\widehat q}$, $\widehat P=e^{\widehat p}$ satisfying $[\widehat q,\widehat p]=i\hbar$.
We then parameterize the $D_5$ quantum curve by
\begin{align}
\frac{\widehat H}{\alpha}
&=\widehat Q\widehat P
-(e_3+e_4)\widehat P
+e_3e_4\widehat Q^{-1}\widehat P
\nonumber\\
&\quad
-(e_1^{-1}+e_2^{-1})\widehat Q
+\frac{E}{\alpha}
-h_2^{-1}e_3e_4(e_5+e_6)\widehat Q^{-1}
\nonumber\\
&\quad
+(e_1e_2)^{-1}\widehat Q\widehat P^{-1}
-h_1(e_1e_2)^{-1}(e_7^{-1}+e_8^{-1})\widehat P^{-1}
+h_1^2(e_1e_2e_7e_8)^{-1}\widehat Q^{-1}\widehat P^{-1},
\label{Hparameter}
\end{align}
where the parameters $h_1,h_2,e_1,\cdots,e_8$ are subject to the constraint
\begin{align}
h_1^2h_2^2=\prod_{i=1}^8e_i.
\label{constraint}
\end{align}
By choosing the parameters suitably, we are able to express the Hamiltonians appearing in the Fredholm determinant \eqref{gpf} for the grand partition functions of the $(2,2)$ model, the $(1,1,1,1)$ model and their rank deformations \cite{MNN,KM} connecting the two models.
Quantum-mechanically, Hamiltonians are defined up to similarity transformations which do not change the spectrum.
Combining two degrees of freedom from similarity transformations $\widehat Q\to A\widehat Q$, $\widehat P\to B\widehat P$ with two degrees of freedom in parameterizing 8 asymptotic points with 10 parameters $h_1,h_2,e_1,e_2,\cdots,e_8$, we can fix the gauge by setting
\begin{align}
e_2=e_4=e_6=e_8=1,\quad
e_7=\frac{h_1^2h_2^2}{e_1e_3e_5},
\label{e2468}
\end{align}
where the last equation comes from the constraint \eqref{constraint}.
Then the $D_5$ Weyl group action can be given unambiguously by
\begin{align}
s_1&:(\tilde h_1,\tilde h_2,e_1,e_3,e_5;\alpha)\mapsto
\biggl(\frac{e_1e_3e_5}{\tilde h_1\tilde h_2^2},\tilde h_2,e_1,e_3,e_5;\alpha\biggr),
\nonumber\\
s_2&:(\tilde h_1,\tilde h_2,e_1,e_3,e_5;\alpha)\mapsto
\biggl(\frac{\tilde h_1}{e_3},h_2,e_1,\frac{1}{e_3},e_5;e_3\alpha\biggr),
\nonumber\\
s_3&:(\tilde h_1,\tilde h_2,e_1,e_3,e_5;\alpha)\mapsto
\biggl(\tilde h_1,\frac{e_1e_5}{\tilde h_1\tilde h_2},e_1,\frac{e_1e_3e_5}{\tilde h_1\tilde h_2^2},e_5;\alpha\biggr),
\nonumber\\
s_4&:(\tilde h_1,\tilde h_2,e_1,e_3,e_5;\alpha)\mapsto
\biggl(\frac{\tilde h_1\tilde h_2}{e_1e_5},\tilde h_2,\frac{\tilde h_2}{e_5},e_3,\frac{\tilde h_2}{e_1};\alpha\biggr),
\nonumber\\
s_5&:(\tilde h_1,\tilde h_2,e_1,e_3,e_5;\alpha)\mapsto
\biggl(\tilde h_1,\frac{\tilde h_2}{e_1},\frac{1}{e_1},e_3,e_5;\frac{\alpha}{e_1}\biggr),
\label{s12345}
\end{align}
where we have defined $\tilde h_1=qh_1$, $\tilde h_2=q^{-1}h_2$ with $q=e^{i\hbar}$.
See figure \ref{dynkin} for our numbering of the simple root for the $D_5$ algebra.

In order to study the Fredholm determinant $\Det(1+z\widehat H^{-1})$ \eqref{gpf}, it is important to extract information from the Hamiltonian $\widehat H$.
The standard process for it is to consider the Schr\"odinger equation for the wave function
\begin{align}
\biggl[\frac{\widehat H}{\alpha}+\frac{z}{\alpha}\biggr]\Psi(x)=0.
\end{align}
Although our supersymmetric gauge theories correspond to special choices of the parameters $(\tilde h_1,\tilde h_2,e_1,e_3,e_5;\alpha)$, we leave them arbitrarily to investigate the group-theoretical structure.
Using the parametrization \eqref{Hparameter} with the gauge fixing condition \eqref{e2468}, we can express the Schr\"odinger equation as\footnote{Similar Schr\"odinger equations and their generalizations appear in \cite{Tm,NRY} in the context of Painlev\'e equations.}
\begin{align}
&\frac{(X-e_3)(X-1)}{X}\Psi[q^{-1}X]\nonumber\\
&+\Bigl[-(e_1^{-1}+1)X+\frac{E}{\alpha}-\frac{e_3(e_5+1)}{h_2X}\Bigr]\Psi[X]\nonumber\\
&+\frac{(X-h_1e_7^{-1})(X-h_1)}{e_1X}\Psi[qX]+\frac{z}{\alpha}\Psi[X]
=0,
\end{align}
with $e_7$ defined in \eqref{e2468}.
Here we have used
\begin{align}
\widehat Q\Psi(x)&=e^{\widehat q}\Psi(x)=e^x\Psi(x)=X\Psi[X],\nonumber\\
\widehat P\Psi(x)&=e^{\widehat p}\Psi(x)=e^{\frac{\hbar}{i}\frac{d}{dx}}\Psi(x)=\Psi(x-i\hbar)=\Psi[q^{-1}X],
\label{QPaction}
\end{align}
where in the last equalities we denote $X=e^x$ and consider $\Psi(x)$ to be a function of $X$ which is denoted by $\Psi[X]$.
We can further introduce the ratio of the wave functions
\begin{align}
P[X]=\frac{\Psi[q^{-1}X]}{\Psi[X]},
\label{vx}
\end{align}
by mimicking the action of the momentum operator $\widehat P$ in \eqref{QPaction} and rewrite the equation as
\begin{align}
&\frac{(X-e_3)(X-1)}{X}P[X]
-(e_1^{-1}+1)X+\frac{E}{\alpha}-\frac{e_3(e_5+1)}{h_2X}
+\frac{(X-h_1e_7^{-1})(X-h_1)}{e_1XP[qX]}
+\frac{z}{\alpha}
=0.
\label{Schrodinger}
\end{align}
Then we can solve this equation order by order in the large $z$ expansion and find
\begin{align}
P[X]=-\frac{z}{\alpha}\frac{X}{(X-e_3)(X-1)}
+\frac{(e_1^{-1}+1)X^2-EX/\alpha+h_2^{-1}e_3(e_5+1)}{(X-e_3)(X-1)}
+O(z^{-1}).
\label{zexpansion}
\end{align}

Classically, the A-periods and the B-periods are defined respectively by integrating the meromorphic one-form $pdx$ along the corresponding cycles.
The definition can be generalized into quantum periods using the wave function or $P[X]$ \eqref{vx} introduced above.
Namely, since the classical A-period is given by
\begin{align}
\oint pdx=\oint\frac{\log P}{X}dX,
\end{align}
using the exponential canonical variables $X=e^x$ and $P=e^p$, it is natural to define the quantum A-period as
\begin{align}
\Pi_A(z)=\Res_{X=0}\frac{1}{X}\log\frac{P[X]}{-zX/\bigl(\alpha(X-e_3)(X-1)\bigr)},
\label{PiA}
\end{align}
by picking up the residue at $X=0$.
Using the quantum A-period, the quantum mirror map is given by
\begin{align}
\log z_\text{eff}=\log z+\Pi_A(z).
\label{zeff}
\end{align}
We shall adopt these setups to study the quantum mirror map for the $D_5$ quantum curve in the next section.

\section{Quantum mirror map in characters}\label{QMPcharacter}

In the previous section we have recapitulated the ideas of quantum curves and quantum mirror maps.
After fully identifying the $D_5$ Weyl group action in the quantum curve, it is natural to expect that the Weyl group action is helpful for studying the quantum mirror map.
At the same time, for the consistency of the definition of the A-period, the mirror map has to be symmetric under the Weyl group action.
In this section, we explicitly perform the analysis and observe several interesting structures.
As in the next section, we can apply these structures to the study of superconformal Chern-Simons theories.

\subsection{Group-theoretical structure}

In this subsection we embark on the study of the quantum mirror map for the $D_5$ curve.
We can perform the large $z$ expansion for the ratio of the wave functions $P[X]$ \eqref{zexpansion} and the A-period $\Pi_A(z)$ \eqref{PiA} order by order.
Here we make several observations for the consistency of the results with the Weyl group action.

Our first observation from the first few orders in the large $z$ expansion is that the quantum A-period
\begin{align}
\Pi_A(z)=\sum_{\ell=1}^\infty z^{-\ell}(\Pi_A)_\ell,
\end{align}
is given by the general expression
\begin{align}
&(\Pi_A)_1=E,\quad
(\Pi_A)_2=-\frac{E^2}{2}-A_2,\quad
(\Pi_A)_3=\frac{E^3}{3}+2EA_2+A_3,\nonumber\\
&(\Pi_A)_4=-\frac{E^4}{4}-3E^2A_2-3EA_3-A_4,\quad
(\Pi_A)_5=\frac{E^5}{5}+4E^3A_2+6E^2A_3+4EA_4+A_5,
\end{align}
where $A_\ell$ is a function of the parameters $q$, $(h_1, h_2, e_1, e_3, e_5)$ and $\alpha$.
Namely, if we expand the coefficients of the A-period $(\Pi_A)_\ell$ for each order $\ell$ in $E$, a higher order term contains lower orders in the nest as in
\begin{align}
\Pi_A(z)
&=\log(1+Ez^{-1})-\sum_{\ell=1}^\infty\frac{(-1)^\ell A_\ell}{z^\ell(1+Ez^{-1})^\ell}.
\end{align}
Note that although for our current case $A_{\ell=1}$ is vanishing, we introduce it from the consistency with other terms.
If we redefine $z$ by
\begin{align}
z_E=z+E,
\label{Eshift}
\end{align}
the quantum mirror map \eqref{zeff} is reexpressed by
\begin{align}
\log z_\text{eff}=\log z_E-\sum_{\ell=1}^\infty\frac{(-1)^\ell A_\ell}{z_E^{\ell}}.
\label{qmm}
\end{align}
The explicit forms of $A_2$ and $A_3$, after fixing the gauge and solving the constraint \eqref{e2468}, are
\begin{align}
\frac{A_2}{\tilde\alpha^2}
&=\frac{e_3}{e_1}+\frac{\tilde h_1}{e_1}+\frac{\tilde h_1e_3}{e_1}+\frac{e_3e_5}{\tilde h_2^2}
+\frac{e_3e_5}{\tilde h_1\tilde h_2^2}+\frac{e_3^2e_5}{\tilde h_1\tilde h_2^2}+\frac{e_3}{\tilde h_2}
+\frac{e_3}{\tilde h_2e_1}+\frac{e_3e_5}{\tilde h_2}+\frac{e_3e_5}{\tilde h_2e_1},\nonumber\\
\frac{A_3}{\tilde\alpha^3}&=(q^{\frac{1}{2}}+q^{-\frac{1}{2}})
\biggl(\frac{\tilde h_1e_3}{e_1^2}+\frac{\tilde h_1e_3}{e_1}
+\frac{e_3^2e_5}{\tilde h_1\tilde h_2^3}+\frac{e_3^2e_5^2}{\tilde h_1\tilde h_2^3}
+\frac{e_3e_5}{\tilde h_2^2}+\frac{e_3e_5}{\tilde h_2^2e_1}
+\frac{e_3^2e_5}{\tilde h_2^2}+\frac{e_3^2e_5}{\tilde h_2^2e_1}\nonumber\\
&\qquad+\frac{e_3^2e_5}{\tilde h_1\tilde h_2^2}+\frac{e_3^2e_5}{\tilde h_1\tilde h_2^2e_1}
+\frac{e_3}{\tilde h_2e_1}+\frac{e_3^2}{\tilde h_2e_1}
+\frac{e_3e_5}{\tilde h_2e_1}+\frac{e_3^2e_5}{\tilde h_2e_1}
+\frac{\tilde h_1e_3}{\tilde h_2e_1}+\frac{\tilde h_1e_3e_5}{\tilde h_2e_1}\biggr),
\end{align}
with $\tilde\alpha$ defined by $\tilde\alpha=\alpha/\sqrt{q}$.

As a second observation, we note that $A_2$ contains $10$ terms while $A_3$, after factoring out $(q^{\frac{1}{2}}+q^{-\frac{1}{2}})$, contains 16 terms which are reminiscent of the characters of the representations ${\bf 10}$ and ${\bf 16}$ or $\overline{\bf 16}$.
Obviously this identification is not quite correct, since the characters for real representations such as ${\bf 10}$ should be symmetric under reversing the powers.
Nevertheless, it is not difficult to observe that the combinations $A_2$ and $A_3$ are invariant under the $D_5$ Weyl group \eqref{s12345} and should be expressed in terms of the $D_5$ characters after suitable modifications.

To identify the results as characters correctly, we need to reconsider the role of the parameter $\alpha$ or $\tilde\alpha$.
In parameterizing the quantum Hamiltonian in \eqref{Hparameter}, the parameter $\alpha$ is simply an overall factor.
On one hand, in identifying the $D_5$ Weyl group action it turns out that this parameter transforms non-trivially as in \eqref{s12345}.
On the other hand, since the transformations \eqref{s12345} on the remaining parameters $(\tilde h_1,\tilde h_2,e_1,e_3,e_5)$ already generate the $D_5$ Weyl group, the parameter $\alpha$ is redundant in the group action.
This means that it should be possible for us to construct a combination from the parameters $(\tilde h_1,\tilde h_2,e_1,e_3,e_5)$ which transforms exactly in the same manner as the parameter $\alpha$.
It turns out that the combination transforming as the parameter $\tilde\alpha=\alpha/\sqrt{q}$ can be constructed explicitly and we identify them by
\begin{align}
\tilde\alpha=(\tilde h_2)^{\frac{1}{2}}e_1^{\frac{1}{4}}e_3^{-\frac{1}{2}}e_5^{-\frac{1}{4}}.
\label{alphahe}
\end{align}
As in section \ref{conclude} the physical meaning of this identification is not very clear to us and needs further clarifications.

To match with the standard characters, we need to switch the fundamental weights identified for the $D_5$ quantum curve in \cite{KMN} into those in the standard orthonormal basis
\begin{align}
\omega_1=(1,-1,0,0,-1)\quad&\leftrightarrow\quad\omega^\perp_1=\biggl(\frac{1}{2},\frac{1}{2},\frac{1}{2},\frac{1}{2},\frac{1}{2}\biggr),\nonumber\\
\omega_2=(1,-1,0,1,-1)\quad&\leftrightarrow\quad\omega^\perp_2=\biggl(\frac{1}{2},\frac{1}{2},\frac{1}{2},\frac{1}{2},-\frac{1}{2}\biggr),\nonumber\\
\omega_3=(1,-2,0,0,-2)\quad&\leftrightarrow\quad\omega^\perp_3=(1,1,1,0,0),\nonumber\\
\omega_4=(0,-1,0,0,-2)\quad&\leftrightarrow\quad\omega^\perp_4=(1,1,0,0,0),\nonumber\\
\omega_5=(0,0,1,0,-1)\quad&\leftrightarrow\quad\omega^\perp_5=(1,0,0,0,0).
\end{align}
Namely, as in \cite{KM}, we express the powers by the fundamental weights of \cite{KMN}, $(\omega_1,\omega_2,\omega_3,\omega_4,\omega_5)$, switch them into those in the orthonormal basis, $(\omega^\perp_1,\omega^\perp_2,\omega^\perp_3,\omega^\perp_4,\omega^\perp_5)$,
\begin{align}
&(\tilde h_1,\tilde h_2,e_1,e_3,e_5)=\tilde h_1^{(1,0,0,0,0)}\tilde h_2^{(0,1,0,0,0)}e_1^{(0,0,1,0,0)}e_3^{(0,0,0,1,0)}e_5^{(0,0,0,0,1)}\nonumber\\
&=(\tilde h_1)^{2\omega_1-\omega_3}
(\tilde h_2)^{2\omega_1-2\omega_3+\omega_4}
e_1^{-\omega_1+\omega_3-\omega_4+\omega_5}
e_3^{-\omega_1+\omega_2}
e_5^{-\omega_1+\omega_3-\omega_4}\nonumber\\
&\leftrightarrow
(\tilde h_1)^{2\omega^\perp_1-\omega^\perp_3}
(\tilde h_2)^{2\omega^\perp_1-2\omega^\perp_3+\omega^\perp_4}
e_1^{-\omega^\perp_1+\omega^\perp_3-\omega^\perp_4+\omega^\perp_5}
e_3^{-\omega^\perp_1+\omega^\perp_2}
e_5^{-\omega^\perp_1+\omega^\perp_3-\omega^\perp_4}\nonumber\\
&=\tilde h_1^{(0,0,0,1,1)}
\tilde h_2^{(0,0,-1,1,1)}
e_1^{(\frac{1}{2},-\frac{1}{2},\frac{1}{2},-\frac{1}{2},-\frac{1}{2})}
e_3^{(0,0,0,0,-1)}
e_5^{(-\frac{1}{2},-\frac{1}{2},\frac{1}{2},-\frac{1}{2},-\frac{1}{2})}\nonumber\\
&=\biggl(\sqrt{\frac{e_1}{e_5}},\frac{1}{\sqrt{e_1e_5}},\frac{\sqrt{e_1e_5}}{\tilde h_2},\frac{\tilde h_1\tilde h_2}{\sqrt{e_1e_5}},\frac{\tilde h_1\tilde h_2}{e_3\sqrt{e_1e_5}}\biggr),
\end{align}
and identify the result as ${\bm q}=(q_1,q_2,q_3,q_4,q_5)$.
This is solved reversely by
\begin{align}
\tilde h_1=q_3q_4,\quad\tilde h_2=\frac{1}{q_2q_3},\quad
e_1=\frac{q_1}{q_2},\quad e_3=\frac{q_4}{q_5},\quad e_5=\frac{1}{q_1q_2}.
\end{align}
With this parameterization $A_\ell$ is now a function of $q$ and ${\bm q}=(q_1,q_2,q_3,q_4,q_5)$.
We then find that
\begin{align}
A_2&=q_1+q_1^{-1}+q_2+q_2^{-1}+q_3+q_3^{-1}+q_4+q_4^{-1}+q_5+q_5^{-1},\nonumber\\
A_3/(q^{\frac{1}{2}}+q^{-\frac{1}{2}})
&=(q_1+q_2+q_3+q_4+q_5
\nonumber\\
&\hspace{-16mm}
+q_1q_2q_3+q_1q_2q_4+q_1q_2q_5+q_1q_3q_4+q_1q_3q_5+q_1q_4q_5+q_2q_3q_4+q_2q_3q_5+q_2q_4q_5+q_3q_4q_5
\nonumber\\
&\hspace{-16mm}+q_1q_2q_3q_4q_5)/\sqrt{q_1q_2q_3q_4q_5},
\end{align}
which are nothing but the characters $\chi_{\bf{10}}({\bm q})$ and $\chi_{\bm{16}}({\bm q})$.
Hence, in this subsection we conclude that the quantum A-period is described by the group-theoretical language of characters if we correctly identify the overall parameter $\tilde\alpha$ by \eqref{alphahe}.

\subsection{Multi-covering structure}\label{MCS}

In the previous subsection, we have found that the results of the quantum A-period are given by the characters.
Exactly the same structure works for the B-period and this further enables us to count the contribution for various representations and summarize the results by multiplicities of representations \cite{MNY}.
We hope to perform the same analysis for the A-period.
If we proceed to higher orders $A_{\ell(\ge 4)}$ in \eqref{qmm}, we find, however, fractions.
In this subsection, we explain how to take care of the fractions by identifying the multi-covering structure correctly and derive the multiplicities of the characters.

\begin{table}[t!]
\begin{align*}
\epsilon'_1&=0,\nonumber\\
\epsilon'_2&=\chi_{\bf 10},\nonumber\\
\epsilon'_3&=(q^{\frac{1}{2}}+q^{-\frac{1}{2}})\chi_{\bf 16},\nonumber\\
\epsilon'_4&=(q^2+q^{-2})\chi_{\bf 1}
+(q+q^{-1})(\chi_{\bf 45}+3\chi_{\bf 1})
+(-\chi_{\bf 54}+\chi_{\bf 45}+3\chi_{\bf 1}),\nonumber\\
\epsilon'_5&=(q^{\frac{5}{2}}+q^{-\frac{5}{2}})\chi_{\overline{\bf 16}}
+(q^{\frac{3}{2}}+q^{-\frac{3}{2}})(\chi_{\overline{\bf 144}}+3\chi_{\overline{\bf 16}})
+(q^{\frac{1}{2}}+q^{-\frac{1}{2}})3\chi_{\overline{\bf 16}},\nonumber\\
\epsilon'_6&=(q^4+q^{-4})\chi_{\bf 10}
+(q^3+q^{-3})(\chi_{\bf 120}+4\chi_{\bf 10})
+(q^2+q^{-2})(\chi_{\bf 320}+\chi_{\overline{\bf 126}}
+3\chi_{\bf 120}+9\chi_{\bf 10})\nonumber\\
&\qquad
+(q+q^{-1})(3\chi_{\bf 120}+8\chi_{\bf 10})
+(\chi_{\bf 320}+2\chi_{\bf 120}+9\chi_{\bf 10}),\nonumber\\
\epsilon'_7&=(q^{\frac{11}{2}}+q^{-\frac{11}{2}})\chi_{\bf 16}
+(q^{\frac{9}{2}}+q^{-\frac{9}{2}})(\chi_{\bf 144}+4\chi_{\bf 16})
+(q^{\frac{7}{2}}+q^{-\frac{7}{2}})(\chi_{\bf 560}+4\chi_{\bf 144}+13\chi_{\bf 16})
\nonumber\\
&\qquad
+(q^{\frac{5}{2}}+q^{-\frac{5}{2}})(\chi_{\bf 720}+4\chi_{\bf 560}+9\chi_{\bf 144}+25\chi_{\bf 16})
+(q^{\frac{3}{2}}+q^{-\frac{3}{2}})(3\chi_{\bf 560}+8\chi_{\bf 144}+27\chi_{\bf 16})
\nonumber\\
&\qquad
+(q^{\frac{1}{2}}+q^{-\frac{1}{2}})(\chi_{\bf 720}+3\chi_{\bf 560}+9\chi_{\bf 144}+27\chi_{\bf 16}),\nonumber\\
\epsilon'_8&=(q^8+q^{-8})\chi_{\bf 1}
+(q^7+q^{-7})(\chi_{\bf 45}+3\chi_{\bf 1})
+(q^6+q^{-6})(\chi_{\bf 210}+\chi_{\bf 54}+4\chi_{\bf 45}+10\chi_{\bf 1})
\nonumber\\
&\qquad
+(q^5+q^{-5})(\chi_{\bf 945}+4\chi_{\bf 210}+4\chi_{\bf 54}+14\chi_{\bf 45}+25\chi_{\bf 1})
\nonumber\\
&\qquad
+(q^4+q^{-4})(\chi_{\overline{\bf 1050}}+4\chi_{\bf 945}+\chi_{\bf 770}+13\chi_{\bf 210}+10\chi_{\bf 54}+35\chi_{\bf 45}+53\chi_{\bf 1})
\nonumber\\
&\qquad
+(q^3+q^{-3})(\chi_{\bf 1386}+4\chi_{\overline{\bf 1050}}+10\chi_{\bf 945}+3\chi_{\bf 770}+25\chi_{\bf 210}+19\chi_{\bf 54}+62\chi_{\bf 45}+84\chi_{\bf 1})
\nonumber\\
&\qquad
+(q^2+q^{-2})(3\chi_{\overline{\bf{1050}}}+9\chi_{\bf 945}+2\chi_{\bf 770}+26\chi_{\bf 210}+18\chi_{\bf 54}+69\chi_{\bf 45}+98\chi_{\bf 1})
\nonumber\\
&\qquad
+(q+q^{-1})(\chi_{\bf 1386}+3\chi_{\overline{\bf 1050}}+10\chi_{\bf 945}+3\chi_{\bf 770}+27\chi_{\bf 210}+22\chi_{\bf 54}+73\chi_{\bf 45}+105\chi_{\bf 1})
\nonumber\\
&\qquad
+(4\chi_{\overline{\bf 1050}}+10\chi_{\bf 945}+2\chi_{\bf 770}+28\chi_{\bf 210}+22\chi_{\bf 54}+72\chi_{\bf 45}+104\chi_{\bf 1}).
\end{align*}
\caption{Components $\epsilon'_d(q,{\bm q})$ appearing in a tentative multi-covering structure \eqref{multicover} for the mirror map of degree $1\le d\le 8$ in terms of the $D_5$ character $\chi_{\bf R}({\bm q})$.}
\label{mirrormap}
\end{table}

In \cite{HMO3} it was known that the inverse quantum mirror map is cleaner than the original quantum mirror map.
For this purpose we solve the quantum mirror map \eqref{qmm} inversely
\begin{align}
\log z_E=\log z_\text{eff}+\sum_{\ell=1}^\infty(-1)^\ell E_\ell z_\text{eff}^{-\ell},
\label{zbarzeff}
\end{align}
where $E_\ell$ is given in terms of $A_\ell$ by
\begin{align}
&E_1=A_1,\quad E_2=A_2-A_1^2,\quad E_3=A_3-3A_2A_1+\frac{3}{2}A_1^3,\nonumber\\
&E_4=A_4-4A_3A_1-2A_2^2+8A_2A_1^2-\frac{8}{3}A_1^4,\nonumber\\
&E_5=A_5-5A_4A_1-5A_3A_2+\frac{25}{2}A_3A_1^2+\frac{25}{2}A_2^2A_1-\frac{125}{6}A_2A_1^3+\frac{125}{24}A_1^5.
\end{align}
Note again that $A_{\ell=1}$ is vanishing for the application to our computation, which is included just for consistency.
Furthermore, following \cite{HMMO} let us tentatively adopt a multi-covering structure for $E_\ell$ as in
\begin{align}
E_\ell=\sum_{n|\ell}\frac{\epsilon'_{\frac{\ell}{n}}(q^n,{\bm q}^n)}{n},
\label{multicover}
\end{align}
where $\sum_{n|\ell}$ stands for the summation taken over all the divisors of $\ell$.
Namely, $E_\ell$ is split into the truly degree $\ell$ contributions $\epsilon'_\ell(q,{\bm q})$ (with $n=1$) and those coming from lower degrees.
Using this multi-covering structure, we can identify the multi-covering components $\epsilon'_d(q,{\bm q})$ as the $D_5$ characters.
We proceed to the analysis of very high degrees and list the multi-covering components $\epsilon'_d(q,{\bm q})$ in table \ref{mirrormap}.
For higher representations appearing in table \ref{mirrormap} we have followed the notation of \cite{Y}.
It is interesting to note that all of the coefficients are given by integers, which implies that we have tentatively identified the multi-covering structure of the quantum mirror map correctly.

\begin{table}[t!]
\begin{align*}
\epsilon_1&=0,\nonumber\\
\epsilon_2&=\chi_{\bf 10},\nonumber\\
\epsilon_3&=(q^{\frac{1}{2}}+q^{-\frac{1}{2}})\chi_{\bf 16},\nonumber\\
\epsilon_4&=(q^2+q^{-2})\chi_{\bf 1}
+(q+q^{-1})(\chi_{\bf 45}+3\chi_{\bf 1})
+4\chi_{\bf 1},\nonumber\\
\epsilon_5&=(q^{\frac{5}{2}}+q^{-\frac{5}{2}})\chi_{\overline{\bf 16}}
+(q^{\frac{3}{2}}+q^{-\frac{3}{2}})(\chi_{\overline{\bf 144}}+3\chi_{\overline{\bf 16}})
+(q^{\frac{1}{2}}+q^{-\frac{1}{2}})3\chi_{\overline{\bf 16}},\nonumber\\
\epsilon_6&=(q^4+q^{-4})\chi_{\bf 10}
+(q^3+q^{-3})(\chi_{\bf 120}+4\chi_{\bf 10})
+(q^2+q^{-2})(\chi_{\bf 320}+\chi_{\overline{\bf 126}}+3\chi_{\bf 120}+9\chi_{\bf 10})\nonumber\\
&\qquad
+(q+q^{-1})(\chi_{\overline{\bf 126}}+2\chi_{\bf 120}+9\chi_{\bf 10})
+(\chi_{\bf 320}+2\chi_{\bf 120}+9\chi_{\bf 10}),\nonumber\\
\epsilon_7&=(q^{\frac{11}{2}}+q^{-\frac{11}{2}})\chi_{\bf 16}
+(q^{\frac{9}{2}}+q^{-\frac{9}{2}})(\chi_{\bf 144}
+4\chi_{\bf 16})+(q^{\frac{7}{2}}+q^{-\frac{7}{2}})(\chi_{\bf 560}+4\chi_{\bf 144}+13\chi_{\bf 16})
\nonumber\\
&\qquad
+(q^{\frac{5}{2}}+q^{-\frac{5}{2}})(\chi_{\bf 720}+4\chi_{\bf 560}+9\chi_{\bf 144}+25\chi_{\bf 16})
+(q^{\frac{3}{2}}+q^{-\frac{3}{2}})(3\chi_{\bf 560}+8\chi_{\bf 144}+27\chi_{\bf 16})
\nonumber\\
&\qquad
+(q^{\frac{1}{2}}+q^{-\frac{1}{2}})(\chi_{\bf 720}+3\chi_{\bf 560}+9\chi_{\bf 144}+27\chi_{\bf 16}),\nonumber\\
\epsilon_8&=(q^8+q^{-8})\chi_{\bf 1}
+(q^7+q^{-7})(\chi_{\bf 45}+3\chi_{\bf 1})
+(q^6+q^{-6})(\chi_{\bf 210}+\chi_{\bf 54}+4\chi_{\bf 45}+10\chi_{\bf 1})
\nonumber\\
&\qquad
+(q^5+q^{-5})(\chi_{\bf 945}+4\chi_{\bf 210}+4\chi_{\bf 54}+14\chi_{\bf 45}+25\chi_{\bf 1})
\nonumber\\
&\qquad
+(q^4+q^{-4})(\chi_{\overline{\bf 1050}}+4\chi_{\bf 945}+\chi_{\bf 770}+13\chi_{\bf 210}+10\chi_{\bf 54}+35\chi_{\bf 45}+54\chi_{\bf 1})
\nonumber\\
&\qquad
+(q^3+q^{-3})(\chi_{\bf 1386}+4\chi_{\overline{\bf 1050}}+10\chi_{\bf 945}+3\chi_{\bf 770}+25\chi_{\bf 210}+19\chi_{\bf 54}+62\chi_{\bf 45}+84\chi_{\bf 1})
\nonumber\\
&\qquad
+(q^2+q^{-2})(3\chi_{\overline{\bf{1050}}}+8\chi_{\bf 945}+3\chi_{\bf 770}+27\chi_{\bf 210}+19\chi_{\bf 54}+68\chi_{\bf 45}+102\chi_{\bf 1})
\nonumber\\
&\qquad
+(q+q^{-1})(\chi_{\bf 1386}+3\chi_{\overline{\bf 1050}}+10\chi_{\bf 945}+3\chi_{\bf 770}+27\chi_{\bf 210}+22\chi_{\bf 54}+73\chi_{\bf 45}+105\chi_{\bf 1})
\nonumber\\
&\qquad
+(4\chi_{\overline{\bf 1050}}+10\chi_{\bf 945}+2\chi_{\bf 770}+28\chi_{\bf 210}+22\chi_{\bf 54}+72\chi_{\bf 45}+108\chi_{\bf 1}).
\end{align*}
\caption{Multi-covering components $\epsilon_d(q,{\bm q})$ for the mirror map of degree $1\le d\le 8$ in terms of the $D_5$ character $\chi_{\bf R}({\bm q})$.}
\label{mirrormapv2}
\end{table}

Let us investigate the integers more carefully by comparing them with the BPS indices.
In \cite{MNY} it was found that, when we identify the BPS indices in \cite{HKP} in terms of representations, the representations appearing in degree $d$ are those in the conjugacy class $d$.
In the current study of the quantum mirror map from the A-period, we continue to confirm the same structure in table \ref{mirrormap}.
Besides, the similarity in the structure between the A-period and the B-period is even stronger.
Namely, in identifying representations for the BPS indices, not all of the representations in the same conjugacy class appear.
For example, in the BPS indices of degree 8, the representations ${\bf 660}$ and ${\bf 1050}$ in conjugacy class $0$ are missing (see table 3 of \cite{MNY}).
We find that almost the same set of the representations appear in table \ref{mirrormap} for the quantum mirror map.
There are only two exceptions for this observation.
The first one is the ${\overline{\bf 16}}$ representation in degree 1, where the contribution is absent for the quantum mirror map.
The second one is the ${\bf 54}$ representation in degree 4, where the ${\bf 54}$ representation is missing in the BPS indices.
Also, it is surprising to note that almost all of the representations in table \ref{mirrormap} have the same sign though only the ${\bf 54}$ representation in degree 4 has an inverse sign.
This may indicate that the first exception may be not so harmful while the second one is more serious.

In order to avoid the ${\bf 54}$ representation in degree 4, let us propose another multi-covering structure by introducing signs,
\begin{align}
E_\ell=\sum_{n|\ell}\frac{(-1)^{n+1}\epsilon_{\frac{\ell}{n}}(q^n,{\bm q}^n)}{n}.
\label{multicoverv2}
\end{align}
We then find that the ${\bf 54}$ representation in degree 4 disappears completely and the representations appearing in the quantum mirror map agree with those appearing in the BPS indices at each degree except for the trivial case of degree 1.
We list the results in table \ref{mirrormapv2}.

As an aside, we comment on the possibility of presenting our results in terms of the $\text{su}(2)$ character $\chi_j(q)=(q^{j+\frac{1}{2}}-q^{-(j+\frac{1}{2})})/(q^{\frac{1}{2}}-q^{-\frac{1}{2}})$.
In this case the absolute values of the integer coefficients are in general smaller while we encounter again some unpleasant minus signs.

\section{Quantum mirror map for matrix models}\label{mm}

In the previous section we have found that the quantum mirror map is given cleanly in terms of the characters.
Now we can apply our general results of the quantum mirror map to super Chern-Simons matrix models in this section.
The superconformal Chern-Simons theories originate from generalizations of the ABJM theory, which describes the worldvolume of M2-branes on the target space ${\mathbb C}^4/{\mathbb Z}_k$.
After applying the localization technique for supersymmetric theories, the partition functions of the superconformal Chern-Simons theories reduce to matrix models, which we call super Chern-Simons matrix models.
The grand potentials for the super Chern-Simons matrix models are given by the free energy of topological strings, after redefining the chemical potentials suitably.
In this section we show that the quantum A-period found in the previous section gives directly the redefinition.

The inverse function of the redefinition of the chemical potential\footnote{In this section we add superscripts $\text{m}=(2,2)\text{ or }(1,1,1,1)$ to coefficients $e_\ell$ in order to specify the model under discussion.
The coefficients $e^\text{m}_\ell$ for the mirror map should not be confused with the parameters of the curve including $e_1,e_2,\cdots, e_8$ in \eqref{Hparameter}.}
\begin{align}
\mu=\mu_\text{eff}+\sum_{\ell=1}^\infty(-1)^\ell e^\text{m}_\ell e^{-\ell\mu_\text{eff}},
\label{mueff}
\end{align}
is given by \cite{MN3}
\begin{align}
&e^{(2,2)}_1=4,\quad
e^{(2,2)}_2=2\cos 2\pi k,\quad
e^{(2,2)}_3=\frac{8}{3}(2+3\cos 2\pi k),\nonumber\\
&e^{(2,2)}_4=16+32\cos 2\pi k+17\cos 4\pi k,\nonumber\\
&e^{(2,2)}_5=\frac{4}{5}(101+200\cos 2\pi k+160\cos 4\pi k+40\cos 6\pi k),
\label{e22}
\end{align}
for the $(2,2)$ matrix model with gauge group U$(N)_k\times$U$(N)_0\times$U$(N)_{-k}\times$U$(N)_0$, while it is given by
\begin{align}
&e^{(1,1,1,1)}_1=4\cos\frac{\pi k}{2},\quad
e^{(1,1,1,1)}_2=2\cos\pi k,\quad
e^{(1,1,1,1)}_3=\frac{8}{3}(2+3\cos\pi k)\cos\frac{3\pi k}{2},\nonumber\\
&e^{(1,1,1,1)}_4=(17+32\cos\pi k+16\cos 2\pi k)\cos 2\pi k,\nonumber\\
&e^{(1,1,1,1)}_5=\frac{4}{5}(101+190\cos\pi k+140\cos 2\pi k+60\cos 3\pi k+10\cos 4\pi k)\cos\frac{5\pi k}{2},
\label{e1111}
\end{align}
for the $(1,1,1,1)$ matrix model with gauge group U$(N)_k\times$U$(N)_{-k}\times$U$(N)_k\times$U$(N)_{-k}$.
For the redefinition for the $(1,1,1,1)$ model, see appendix \ref{1111} following the method of \cite{HM}.

In \cite{MN3} the redefinition was further decomposed into multi-covering components.
However, due to the constant shift in \eqref{Eshift}, the multi-covering structure should be modified.
For this reason, we list the redefinition in \eqref{e22} and \eqref{e1111} without referring to the multi-covering structure.
Accordingly, to explain the redefinition using the quantum mirror map we identify the chemical potential $\mu$ and the effective one $\mu_\text{eff}$ as $\log z$ and $\log z_\text{eff}$ respectively and express $\log z$ (instead of $\log z_E$) in terms of $\log z_\text{eff}$.
For this purpose we first rewrite \eqref{Eshift} as\footnote{We add superscripts $\text{m}$ to $E$ and $E_\ell$ as well to specify the model.}
\begin{align}
\log z=\log z_E+\log(1-E^\text{m}z_E^{-1}),
\end{align}
and substitute \eqref{zbarzeff} in it to find
\begin{align}
\log z=\log z_\text{eff}+\sum_{\ell=1}^\infty(-1)^\ell E^\text{m}_\ell z_\text{eff}^{-\ell}
+\log\Bigl(1-E^\text{m}z_\text{eff}^{-1}e^{-\sum_{\ell=1}^\infty(-1)^\ell E^\text{m}_\ell z_\text{eff}^{-\ell}}\Bigr).
\label{zzeff}
\end{align}

For the models the constant shifts $E^\text{m}$ are obtained respectively as
\begin{align}
E^{(2,2)}=4,\quad
E^{(1,1,1,1)}=2(q^{\frac{1}{4}}+q^{-\frac{1}{4}}),
\end{align}
from the spectral operators
\begin{align}
\widehat H^{(2,2)}
=\widehat{\cal Q}^2\widehat{\cal P}^2
=\sum_{\{\pm\}^4}\widehat Q^{\pm\frac{1}{2}}\widehat Q^{\pm\frac{1}{2}}\widehat P^{\pm\frac{1}{2}}\widehat P^{\pm\frac{1}{2}},\quad
\widehat H^{(1,1,1,1)}
=\widehat{\cal Q}\widehat{\cal P}\widehat{\cal Q}\widehat{\cal P}
=\sum_{\{\pm\}^4}\widehat Q^{\pm\frac{1}{2}}\widehat P^{\pm\frac{1}{2}}\widehat Q^{\pm\frac{1}{2}}\widehat P^{\pm\frac{1}{2}},
\end{align}
by picking up relevant constant terms where the total powers are all canceled.
Note that if the reciprocal operators are not located next to each other the constant shift $E^\text{m}$ can be non-trivial due to the commutation relation ${\cal P}^\alpha{\cal Q}^\beta=q^{-\alpha\beta}{\cal Q}^\beta{\cal P}^\alpha$.
Also, as in \cite{MNY,KMN,KM} the U$(1)$ charges of the $D_5$ characters
\begin{align}
{\bm q}=(q_1,q_2,q_3,q_4,q_5)=(1,e,h^{-1},e,1),
\end{align}
for the two models have to be identified as
\begin{align}
(h,e)^{(2,2)}=(e^{-2\pi ik},1),\quad
(h,e)^{(1,1,1,1)}=(e^{-\pi ik},e^{\pi ik}).
\end{align}
Then, we can substitute the U$(1)$ charges subsequently into the characters listed in \cite{MNY,KM}, into the multi-covering component $\epsilon_d(q,{\bm q})$ in table \ref{mirrormapv2}, into the expression of the multi-covering structure $E^\text{m}_\ell$ \eqref{multicoverv2}, and then into \eqref{zzeff}.
We find that the expressions \eqref{e22} and \eqref{e1111} are reproduced correctly from the substitutions.

\section{Conclusions and discussions}\label{conclude}

In this paper, we have carefully studied the quantum mirror map for the $D_5$ quantum curve.
We have found that, as in the ABJM case, the quantum mirror map is computed from the A-period.
After the computation we find that the results are summarized cleanly in terms of the $D_5$ characters.
Also we have clarified the multi-covering structure for the results.
The coefficients of the characters are given with integers, which justifies strongly our assumption of the multi-covering structure.
These structures enable us to reproduce the redefinition of the chemical potential for two super Chern-Simons matrix models correctly.
In the following we list some further directions we would like to pursue in the future.

Firstly, to identify the results of the quantum A-periods in terms of characters, in \eqref{alphahe} we have identified the overall factor $\alpha$ for the quantum Hamiltonian \eqref{Hparameter}, which is redundant in generating the $D_5$ Weyl group, with a combination of $(\tilde h_1,\tilde h_2,e_1,e_3,e_5)$ transforming in the same manner.
Technically both the quantum curve and the quantum A-periods are invariant under the transformations \eqref{s12345} only when we transform $\alpha$ appropriately.
Hence in presenting the results containing the parameter $\alpha$ in terms of the $D_5$ characters, we cannot neglect its transformation and the easiest way to take care of the transformation for $\alpha$ is to identity it with the remaining parameters.
The physical meaning of the identification is, however, unclear to us.
We would like to understand it more carefully such as from the affine structure of the Weyl group for the Lie algebra in behind.

Secondly, signs of the multi-covering structure need further clarifications.
In the first attempt of the multi-covering structure in \eqref{multicover} we encounter uncomfortable behavior in table \ref{mirrormap} where the representation ${\bf 54}$ appears in degree 4 with a negative integer coefficient.
After modifying signs of the multi-covering structure in \eqref{multicoverv2} we obtain an expression in table \ref{mirrormapv2} where the representation ${\bf 54}$ disappears and the multi-covering component of each degree contains the same set of representations with positive integer coefficients as in the case of the BPS indices.
Due to this reason, we believe that we have correctly identified the multi-covering structure for the quantum mirror map, though the physical meaning of the signs is still unclear to us.

Thirdly, related to the above discussion, the appearance of the identical set of the representations in the mirror maps and the BPS indices is surprising to us.
Probably the admissible representations are encoded purely in the curve itself and do not depend on the cycles of the integration periods.
Indeed, in \cite{HKP} the A-period and the B-period for the classical curve were calculated from the Weierstrass normal form of the curve, where the coefficients of weight $d$ are given by representations in the conjugacy class $d$.
Hence it is not so mysterious that both the mirror maps and the BPS indices of degree $d$ obtained from the two periods are given by those representations as well.
Furthermore, one may naturally expect that both of them should be expressed exactly by the same set of the representations.
To make this statement rigorous, however, we need to first understand the multi-covering structure for the mirror maps since the set of the representations can differ as stated above.
Besides, we also need to clarify how the Weierstrass normal form of the classical curves is lifted for the quantum curves.

Fourthly, aside from our application to the super Chern-Simons matrix models, it was also proposed \cite{AKV} that the classical limit of the quantum mirror map counts BPS states in the presence of D-brane domain walls.
We believe that the interpretation is consistent with our result in table \ref{mirrormapv2}, which implies that, after reducing the mirror map to the classical limit by setting $q\to 1$, the coefficients are all positive integers.
By comparing this interpretation with the super Chern-Simons matrix models discussed here, this may suggest that the worldsheet instantons are accompanied by an infinite tower of BPS states and the set of positive integers obtained from the quantum mirror map counts these states.
It is an interesting future direction to further pursue the interpretation of the quantum mirror map in the BPS state counting.
Especially, we would like to know the geometrical interpretation of our quantum deformation parameter $q$.

Fifthly, compared with the BPS indices (see the tables in \cite{MNY}), the multiplicities in table \ref{mirrormapv2} are much larger, which may suggest a further structure to be explored.
If we rewrite the results in terms of the $\text{su}(2)$ characters, the multiplicities become smaller, though now the unpleasant minus signs appear again. 
This may imply a more intricate spin dependence.
For example, although the BPS indices are classified by the spins $(j_\text{L},j_\text{R})$, only one $\text{SU}(2)$ spin is taken into considerations for the quantum mirror map.

Sixthly, we have investigated the mirror map responsible for the redefinition of the chemical potential in the super Chern-Simons matrix models in this paper.
There are, however, some properties which seem simpler in terms of the original chemical potential $\mu$.
For example, if we take a close look at the coefficients of the grand potential (as in appendix A of \cite{HMO3}) apparently there are some relations between $k$ and its double.
Also in \cite{MS1} the orientifold projection of the ABJM matrix model was studied and it was found that the extra terms with half powers are summarized in a simple function (see (2.45) in \cite{MS1}).
These relations are not so clear after the redefinition of the chemical potential and we would like to understand the role of the redefinition.

Finally, it was known that the ABJM matrix model enjoys the integrability of the $q$-Painlev\'e equation \cite{BGT} and the 2-dimensional Toda lattice hierarchy \cite{MaMo,FM1,FM2}.
It is interesting to investigate how the integrability works for the $D_5$ curve and how the integrability imposes constraints on the quantum periods.
The similarity of the group-theoretical structure and the multi-covering structure between the BPS indices and the quantum mirror map as we have discussed above may also be related to the integrability.

\appendix

\section{$(1,1,1,1)$ model without rank deformations}\label{1111}

Before going to the $(1,1,1,1)$ model, let us focus on the ABJM model which is the $(1,1)$ model.
The grand potential of the ABJM model $J(\mu)=J^{(1,1)}(\mu)$ is given in terms of the effective chemical potential $\mu_\text{eff}$ formally as
\begin{align}
J(\mu)=F(\mu_\text{eff}),
\end{align}
where $F(\mu_\text{eff})$ is a fixed function.
Here the effective chemical potential $\mu_\text{eff}$ is related to the original chemical potential $\mu$ by \eqref{mueffredef}
\begin{align}
\mu_\text{eff}=\mu+\frac{\pi^2k}{2}\sum_{\ell=1}^\infty a_\ell e^{-2\ell\mu},
\end{align}
with
\begin{align}
&a_1=-\frac{4}{\pi^2k}\cos\frac{\pi k}{2},\quad
a_2=-\frac{2}{\pi^2k}(4+5\cos\pi k),\nonumber\\
&a_3=-\frac{8}{3\pi^2k}\cos\frac{\pi k}{2}(19+28\cos\pi k+3\cos 2\pi k),\nonumber\\
&a_4=-\frac{1}{\pi^2k}(364+560\cos\pi k+245\cos 2\pi k+48\cos 3\pi k+8\cos 4\pi k),\nonumber\\
&a_5=-\frac{8}{5\pi^2k}\cos\frac{\pi k}{2}
(2113+3374\cos\pi k+1751\cos 2\pi k+525\cos 3\pi k\nonumber\\
&\qquad\qquad\qquad+145\cos 4\pi k+25\cos 5\pi k+5\cos 6\pi k).
\label{aell}
\end{align}

In \cite{HM} it was derived that the grand potential of the $(1,1,1,1)$ model $\widetilde J(\mu)=J^{(1,1,1,1)}(\mu)$ is given in terms of that of the ABJM model $J(\mu)=J^{(1,1)}(\mu)$ as
\begin{align}
e^{\widetilde J(\mu)}=\sum_{n=-\infty}^\infty
e^{J(\frac{\mu+\pi i}{2}+2\pi in)+J(\frac{\mu-\pi i}{2}-2\pi in)},
\end{align}
since the $(1,1,1,1)$ model is obtained by repeating the quiver of the ABJM model twice.
Motivated by this relation, let us define the quantum mirror map for the $(1,1,1,1)$ model as
\begin{align}
\widetilde\mu_\text{eff}=\mu+\pi^2k(-a_1e^{-\mu}+a_2e^{-2\mu}-a_3e^{-3\mu}+\cdots),
\label{qmm1111}
\end{align}
using the same coefficient $a_\ell$ \eqref{aell} appearing in the case of the ABJM model.
Then, the relation derived from it
\begin{align}
&\frac{\widetilde\mu_\text{eff}\pm\pi i}{2}\pm 2\pi in
=\frac{\mu\pm\pi i}{2}\pm 2\pi in\nonumber\\
&\qquad+\frac{\pi^2k}{2}
(a_1e^{-2(\frac{\mu\pm\pi i}{2}\pm 2\pi in)}+a_2e^{-4(\frac{\mu\pm\pi i}{2}\pm 2\pi in)}+a_3e^{-6(\frac{\mu\pm\pi i}{2}\pm 2\pi in)}+\cdots),
\end{align}
implies that the grand potential of the $(1,1,1,1)$ model $\widetilde J(\mu)=\widetilde F(\widetilde\mu_\text{eff})$ expressed in terms of the effective chemical potential $\widetilde\mu_\text{eff}$ is given by
\begin{align}
e^{\widetilde F(\widetilde\mu_\text{eff})}=\sum_{n=-\infty}^\infty
e^{F(\frac{\widetilde\mu_\text{eff}+\pi i}{2}+2\pi in)+F(\frac{\widetilde\mu_\text{eff}-\pi i}{2}-2\pi in)}.
\end{align}
This should justify the definition of the quantum mirror map for the $(1,1,1,1)$ model \eqref{qmm1111}.
In section \ref{mm} we only discuss the $(1,1,1,1)$ model and we refer to $\widetilde\mu_\text{eff}$ simply as $\mu_\text{eff}$.
By solving reversely we find \eqref{mueff} with \eqref{e1111}.

\section*{Acknowledgements}
We are grateful to Yasuyuki Hatsuda, Hirotaka Hayashi, Naotaka Kubo, Tomoki Nosaka and Yasuhiko Yamada for valuable discussions.
The work of T.F.\ and S.M.\ is supported respectively by Grant-in-Aid for JSPS Fellows \#20J15045 and Grant-in-Aid for Scientific Research (C) \#19K03829.

\end{document}